\documentclass[reprint,
superscriptaddress,
 amsmath,amssymb,
 aps,
]{revtex4-2}

\usepackage{graphicx}
\usepackage{dcolumn}
\usepackage{bm}
\usepackage{xcolor}
\usepackage{placeins}
\usepackage{tikz}
\newcommand{\tikzcircle}[2][black,fill=black]{\tikz[baseline=-0.5ex]\draw[#1,radius=#2] (0,0) circle ;}%
\usepackage{cleveref}

\begin{document}


\title{Highly Tunable Ru-dimer Molecular Orbital State in 6H-perovskite Ba$_3$MRu$_2$O$_9$}

\author{Bo Yuan}
\thanks{These authors contributed equally to this work}
\address{Department of Physics and Astronomy, McMaster University, Hamilton, Ontario L8S 4M1, Canada}
\author{Beom Hyun Kim}
\thanks{These authors contributed equally to this work}
\address{Center for Theoretical Physics of Complex Systems, Institute for Basic Science, Daejeon 34126, Republic of Korea}
\author{Qiang Chen}
\address{Department of Physics and Astronomy, McMaster University, Hamilton, Ontario L8S 4M1, Canada}
\author{Daniel Dobrowolski}
\address{Department of Physics and Astronomy, McMaster University, Hamilton, Ontario L8S 4M1, Canada}
\author{Monika Azmanska}
\address{Department of Physics and Astronomy, McMaster University, Hamilton, Ontario L8S 4M1, Canada}
\author{G.M. Luke}
\address{Department of Physics and Astronomy, McMaster University, Hamilton, Ontario L8S 4M1, Canada}
\author{Shiyu Fan}
\address{National Synchrotron Light Source II, Brookhaven National Laboratory,Upton, NY, 11973, USA}
\author{Valentina Bisogni}
\address{National Synchrotron Light Source II, Brookhaven National Laboratory,Upton, NY, 11973, USA}
\author{Jonathan Pelliciari}
\address{National Synchrotron Light Source II, Brookhaven National Laboratory,Upton, NY, 11973, USA}
\author{J.P. Clancy}
\address{Department of Physics and Astronomy, McMaster University, Hamilton, Ontario L8S 4M1, Canada}
\date{\today}
\begin{abstract}
Molecular orbital (MO) systems with clusters of heavy transition metal (TM) ions are one of the most important classes of model materials for studying the interplay between local physics and effects of itinerancy. Despite a large number of candidates identified in the family of 4d TM materials, an understanding of their physics from competing \textit{microscopic} energy scales is still missing. We bridge this gap by reporting the first resonant inelastic X-ray scattering (RIXS) measurement on a well-known series of Ru dimer systems with a 6H-perovskite structure, Ba$_3$MRu$_2$O$_9$ (M$^{3+}$=In$^{3+}$, Y$^{3+}$, La$^{3+}$). Our RIXS measurements reveal an extremely fragile MO state in these Ru dimer compounds, evidenced by an abrupt change in the RIXS spectrum accompanying a tiny change in the local structure tuned by the M-site ion. By modelling the RIXS spectra, we attribute the enhanced electronic instability in Ba$_3$MRu$_2$O$_9$ to the combined effect of a large hopping and a small spin-orbit coupling in the Ru dimers. The unique combination of energy scales uncovered in the present study make Ru MO systems ideal model systems for studying quantum phase transitions with molecular orbitals.  
\end{abstract}
\maketitle

``Cluster Mott insulators", or materials comprised of ``molecules" of transition metal (TM) ions have been one of the most important classes of model systems in the study of strongly correlated phenomena. Due to strong inter-site hopping that is comparable or even larger than intra-atomic interactions, orbitals from all sites within each ``molecule" or cluster are strongly hybridized to form so-called ``molecular orbitals" (MO). Here, the competition between hopping and intra-atomic interactions, combined with an increase in orbital degeneracy provide the minimal ingredients for many interesting phenomena such as orbital-selective localization \citep{StreltsovPNAS2016,StreltsovPRB2014} and non-local quantum entanglement\citep{ren2024witnessing} that have no analogue in isolated TM ions. Moreover, the exactly solvable nature of the finite-sized problem allows detailed comparison between theory and experiment. These two properties place the MO materials halfway between the extreme limits of a solid and an isolated ion, making them unique model systems for studying the interplay between local (e.g. strong correlation) and non-local physics (e.g. itinerancy). 

\begin{figure}[tb]
\includegraphics[width=0.5\textwidth]{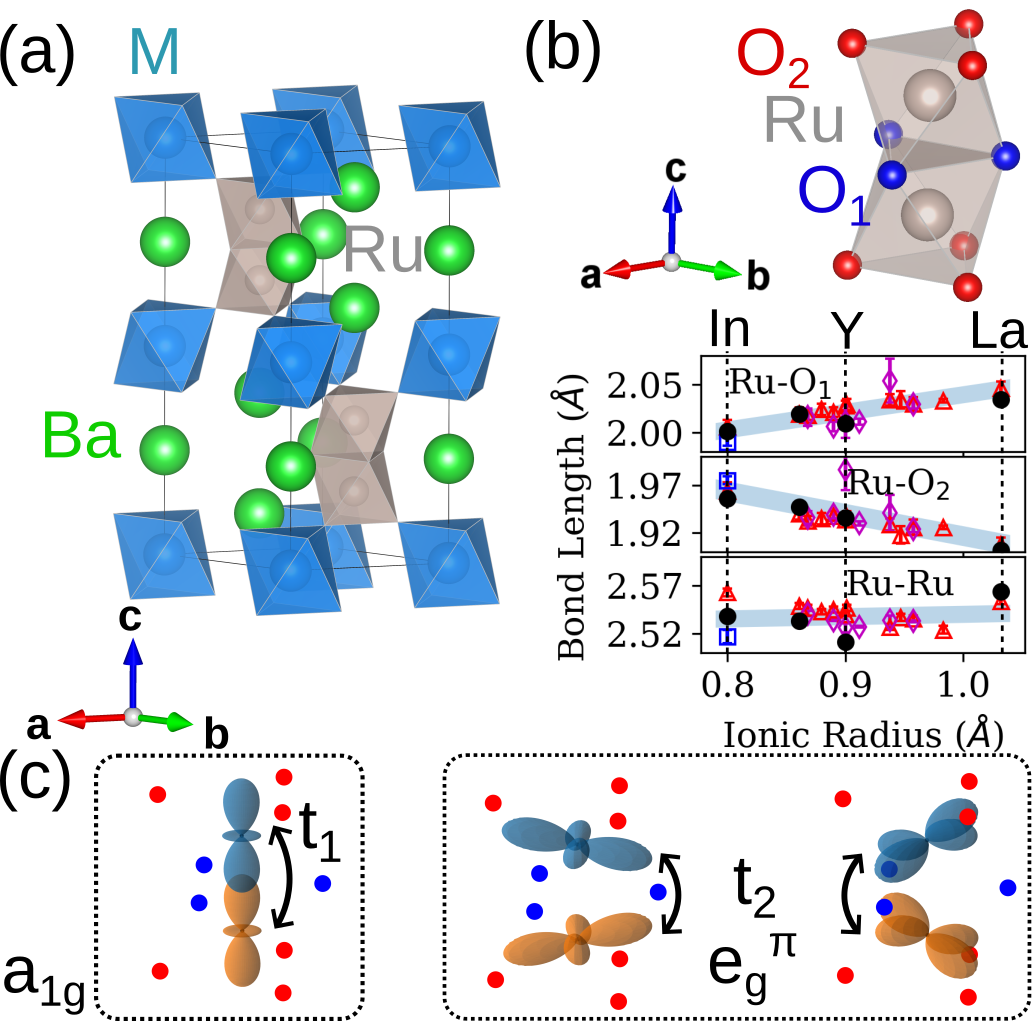}
\caption{\label{structure} (a) Crystal structure of Ba$_3$MRu$_2$O$_9$ without the oxygen ions. (b) Local structure of the Ru$_2$O$_9$ dimer as a function of the ionic radius of the M$^{3+}$ ion. Data points with different symbols are taken from: Ref.~\citep{DOI2001113,DOI2002317,B111504A}(\textcolor{red}{$\bigtriangleup$}), Ref.~\citep{Muller1996,RATH1994119}(\textcolor{magenta}{$\Diamond$}), Ref.~\citep{RIJSSENBEEK199965}(\textcolor{blue}{$\square$}) and Ref.~\citep{PhysRevB.95.184424,PhysRevMaterials.4.064409}(\tikzcircle{2pt}), the last of which used the same sample as the present study. The thick blue trend line in each panel is obtained by a linear fit to all reported bond lengths. The local structural parameters for M$^{3+}$ = In$^{3+}$, Y$^{3+}$ and La$^{3+}$ are denoted by vertical dotted lines. (c) Schematic illustration of the non-degenerate $a_{1g}$ and doubly degenerate $e_g^\pi$ orbitals in the $t_{2g}$ manifold under a trigonal crystal field. The hoppings between them are denoted as $t_1$ and $t_2$, respectively}
\end{figure}

Most MO candidates have been identified from the family of heavy TM materials with 4d or 5d ions such as Ru and Ir, where inter-site hopping is enhanced by the extended nature of the 4d and 5d orbitals \footnote{ Given the large number of 4d/5d MO candidate materials, we cite only a few review articles. For example, see Ref.~\citep{Khmoskiireview2021,Streltsov_2017,Bobcavareview2021}}. A particularly important class of materials are the 6H-perovskites with chemical formula Ba$_3$MB$_2$O$_9$ \citep{SAKAMOTO20062595,DOI2002317,B111504A,Doi_2004jpcm,10.1246/bcsj.76.1093,DOI20043239}, where M is a non-magnetic ion and B=Ir/Ru is magnetic. As shown in Fig.~\ref{structure}, the magnetic unit is a simple face-sharing Ru or Ir dimer with an extremely short Ru-Ru/Ir-Ir bond that favours the formation of MO's. Simplicity and ease of tuning local structure via M-site substitution make these materials ideal platforms for studying MO physics. 

Despite a wealth of experimental evidence for magnetism driven by MO physics in these dimer compounds\citep{PhysRevMaterials.4.064409,PhysRevB.95.184424,doi:10.1021/jacs.9b03389,PhysRevB.91.235147,PhysRevB.96.014434,PhysRevB.97.064408,PhysRevLett.116.097205}, an understanding of their microscopic physics, as well as that of other MO candidates with more complex cluster geometry\citep{Jeong2017,JeongPRB2021,magnaterra2023quasimolecular} became available only recently from resonant inelastic X-ray scattering (RIXS). Since the first measurement on Ba$_3$CeIr$_2$O$_9$\citep{Revelli2019}, Ir L$_3$-edge RIXS has been applied to a large number of Ir dimer systems\citep{CaZnMgSrdimerRIXS,Ti_dimer,In_Irdimer,Aldimer,Aldimer2,ren2024witnessing}, allowing a direct determination of their microscopic electronic Hamiltonian. In contrast, due to the difficulty in conducting analogous RIXS measurements at the Ru L$_3$ edges, direct measurements of the electronic excitations in 4d Ru MO systems is completely missing. The combination of smaller inter-site hopping due to more localized 4d orbitals, and drastically different intra-atomic interactions, such as the significantly reduced 4d spin-orbit coupling (SOC), has the potential to realize qualitatively different microscopic physics in Ru-based MO systems compared to their Ir counterparts, unlocking completely new regimes of MO physics.
  
Here, we address this outstanding question by reporting Ru M$_3$-edge RIXS measurements of the dimer electronic excitations in a well-known series of Ru 6H-perovskites. The choice of Ru M$_3$-edge is motivated by recent works demonstrating the feasibility of Ru M$_3$-edge RIXS \citep{Lebert_2020, RuO2RIXS, Blairnonlocal}, which, compared to the more traditional Ru L$_3$-edge RIXS, uses soft (E$_i$=463~eV) rather than tender X-ray photons (E$_i$=2838~eV) and is therefore technically less challenging. The 6H-perovskite series studied here, Ba$_3$MRu$_2$O$_9$, with M$^{3+}$=In$^{3+}$, Y$^{3+}$, La$^{3+}$ and 7 electrons in the Ru$_2$O$_9$ dimer, has been intensively studied in the past due to its anomalous magnetic properties\citep{PhysRevMaterials.4.064409,PhysRevB.95.184424,DOI2002317} (See Supplemental Materials \citep{supp} for materials characterization and experimental details). 

Our measurements reveal RIXS spectra in the Ba$_3$MRu$_2$O$_9$ series that are qualitatively different from the predictions of a local model, and place them well in the MO limit with a large inter-site hopping. Remarkably, the measured RIXS spectra show an abrupt change with only a tiny change in local structure from Ba$_3$InRu$_2$O$_9$/Ba$_3$YRu$_2$O$_9$ to Ba$_3$LaRu$_2$O$_9$ [at most $1\%-2\%$ change in Ru-O and Ru-Ru bond lengths - see Fig.~\ref{structure}~(b)], thus providing direct evidence for the fragility of electronic states in these Ru dimer compounds. Our theoretical modelling attributes the enhanced electronic instability in the Ru dimer to a large inter-site hopping and reduced SOC compared to its Ir counterparts. A direct manifestation of such instability is the prediction of an electronic transition driven by a very small change in the trigonal distortion of the dimer, accompanied by an abrupt change in the simulated RIXS spectra which is in excellent agreement with the experiment. The high degree of tunability for the electronic states of the Ru dimer uncovered in the present study shows that Ru MO compounds are ideal candidates for studying quantum phase transitions involving molecular orbitals.    

\begin{figure*}[tb]
\includegraphics[width=1\textwidth]{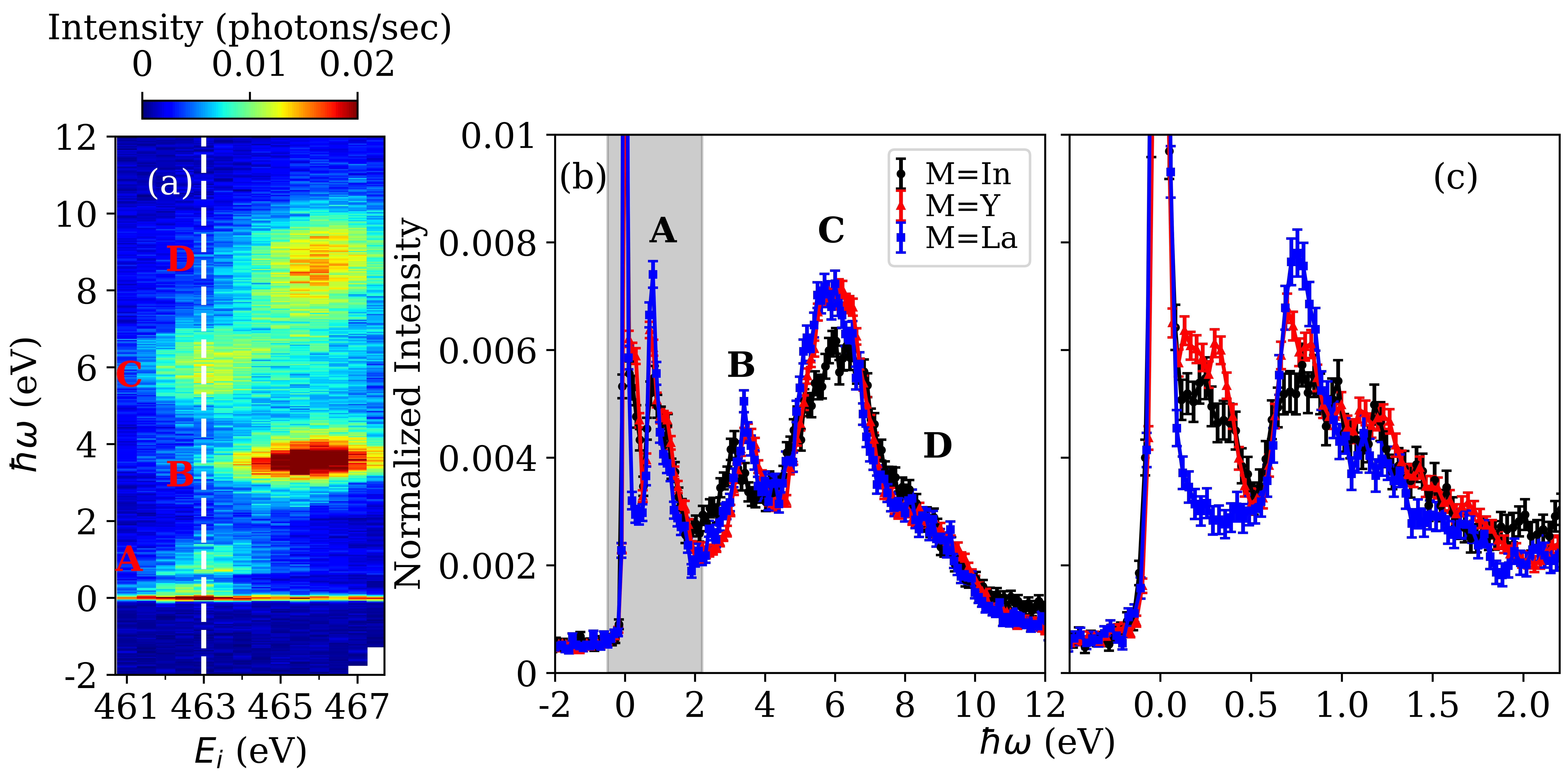}
\caption{\label{rixsdata} (a) Map of RIXS intensity as a function of incident energy, $\mathrm{E_i}$, and energy transfer, $\hbar\omega$, for Ba$_3$YRu$_2$O$_9$. The four features in the RIXS map denote intra-$t_{2g}$ excitations (A), $t_{2g}-e_g$ excitations (B) and charge transfer excitation between the oxygen $p$-orbital and the Ru-$t_{2g}$ (C) and $e_g$ orbitals (D). (b) RIXS intensity as a function of $\hbar\omega$ for Ba$_3$MRu$_2$O$_9$ (M=In, Y, La) using $\mathrm{E_i}$=463~eV. $\mathrm{E_i}$=463~eV, marked by the vertical dashed line in (a), is chosen to resonate with the intra-$t_{2g}$ features [A in (a)]. To directly compare RIXS scans of different compounds, the RIXS intensity is normalized with respect to the integrated intensity between 2~eV and 12~eV covering features B-D, which should be relatively material-independent. (c) RIXS scans focussing on the intra-$t_{2g}$ excitations below $\sim$ 2eV obtained by zooming into the grey shaded region in (b). All measurements shown here are carried out at 40~K.}
\end{figure*}

The overall RIXS features of a Ru dimer are shown by the color map in Fig.~\ref{rixsdata}~(a), where we plot RIXS intensity as a function of incident photon energy, $\mathrm{E_i}$ (horizontal axis), and energy transfer, $\hbar\omega$ (vertical axis) for Ba$_3$YRu$_2$O$_9$. Four excitations can be clearly resolved at (A) $\hbar\omega\lesssim2~$eV, (B) $\hbar\omega\sim$3.5~eV, (C) $\hbar\omega\sim$6~eV and (D) $\hbar\omega\sim$9~eV. Aside from a difference in the absolute $\mathrm{E_i}$'s used in our experiment, corresponding to the $3p_{\frac{3}{2}}\rightarrow 4d$ transition, the main RIXS features shown in Fig.~\ref{rixsdata}~(a) are qualitatively similar to those observed in Ru L$_3$-edge RIXS measurements \citep{Ca2RuO4RIXS,Ca3Ru2O7RIXS,Sr2RuO4RIXS}, allowing us to assign A-D as (A) intra-$t_{2g}$ excitations, (B) $t_{2g}-e_g$ excitations, and charge transfer excitations from the oxygen $p$ orbitals to (C) the Ru $t_{2g}$ and (D) $e_g$ orbitals, respectively. The difference in resonant $\mathrm{E_i}$'s of these features ($\sim$3~eV) corresponds to the different intermediate states where the core electron is excited to the $t_{2g}$ and $e_g$ orbitals for A/C and B/D, respectively. RIXS spectra for Ba$_3$MRu$_2$O$_9$ with different M-site ions are compared in Fig.~\ref{rixsdata}(b) and (c). Fig.~\ref{rixsdata}~(c) is an enlarged version of (b) that focusses on the intra $t_{2g}$ excitations with $\hbar\omega\lesssim$~2eV. The $\mathrm{E_i}$ for these scans is chosen to resonate with the intra-$t_{2g}$ features. As expected from the small changes in local structure going from M=In to M=La, the high energy electronic excitations (B,C,D) are qualitatively similar for the three Ba$_3$MRu$_2$O$_9$ compounds, consistent with minimal changes in the dominant energy scales such as $t_{2g}-e_g$ splitting and Ru-O charge transfer energy across the Ba$_3$MRu$_2$O$_9$ series. 

In contrast to the high energy region, the low energy RIXS spectra of the intra-$t_{2g}$ excitations [Fig.~\ref{rixsdata}~(c)] are drastically different in these three compounds. Two broad peaks can be resolved in Ba$_3$InRu$_2$O$_9$ and Ba$_3$YRu$_2$O$_9$ at $\hbar\omega\sim$0.3~eV and $\hbar\omega\sim$0.7~eV, respectively. Strikingly, these features are very different in Ba$_3$LaRu$_2$O$_9$: the low energy peak is suppressed, while a sharp peak-like feature gains intensity at $\sim$0.7~eV. As we show in the Supplemental Materials\citep{supp}, the suppression of the low energy RIXS intensity below $\hbar\omega<$0.5~eV relative to the high energy feature in Ba$_3$LaRu$_2$O$_9$ is a robust observation independent of $\mathrm{E_i}$'s. 

Qualitatively different RIXS spectra shown in Fig.~\ref{rixsdata}~(c) provide direct evidence for distinct electronic states in Ba$_3$LaRu$_2$O$_9$ and Ba$_3$InRu$_2$O$_9$/Ba$_3$YRu$_2$O$_9$. This is quite surprising considering the small magnitude of the structural change across the Ba$_3$MRu$_2$O$_9$ series [Fig.~\ref{structure}~(b)]. Our observations are in stark contrast to analogous RIXS measurements on face sharing Ir-dimers (Ir$^{5+}$:Ref~\citep{CaZnMgSrdimerRIXS}, Ir$^{4.5+}$:Ref~\citep{Aldimer,Aldimer2,In_Irdimer}, Ir$^{4+}$:Ref~\citep{Ti_dimer,Revelli2019}). There, qualitatively similar RIXS spectra robust against small changes in local structures are observed for compounds with identical Ir valences. A modification of the RIXS spectra by a change in local structure is usually attributed to a change in energies of the excited states. Here, this explanation requires one to move up a bulk of the low-lying excited states at $\hbar\omega\sim0.3~$eV in Ba$_3$InRu$_2$O$_9$ and Ba$_3$YRu$_2$O$_9$ to $\hbar\omega\sim0.7$~eV in Ba$_3$LaRu$_2$O$_9$. However, such a large shift in the excited states seems highly unlikely given at most a $1\%-2\%$ change in Ru-O and Ru-Ru bond lengths from  Ba$_3$YRu$_2$O$_9$ to  Ba$_3$LaRu$_2$O$_9$ [Fig.~\ref{structure}~(b)] with no change in local symmetry\citep{PhysRevMaterials.4.064409,PhysRevB.95.184424,DOI2002317}.   

\begin{figure*}[tb]
\includegraphics[width=1\textwidth]{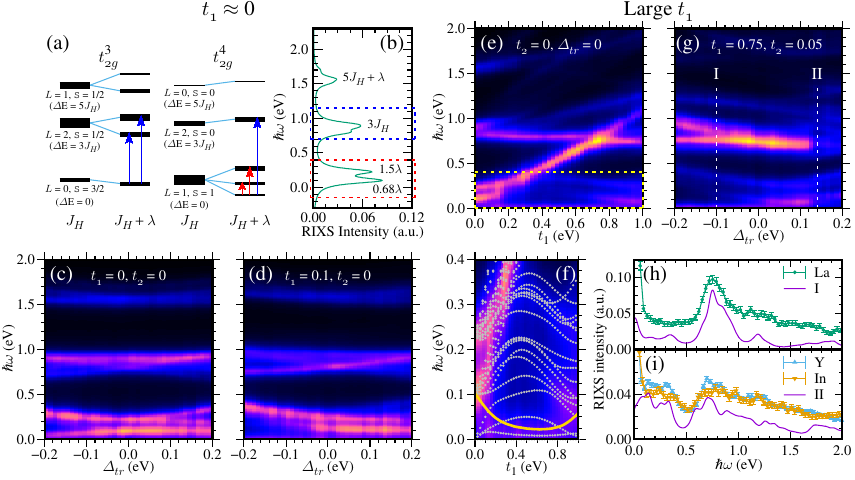}
\caption{\label{theory1} Energy levels and simulated RIXS spectra \textbf{(a-d)} in the localized limit with $t_1\approx$0 and \textbf{(e-i)} in the limit of large $t_1$. \textbf{(a-d):} (a) The energy level diagram for the $t_{2g}^3$ and $t_{2g}^4$ sites in a Ru$_2$O$_9$ dimer with 7 electrons (the degeneracy of each multiplet is proportional to the line thickness) and (b) the corresponding RIXS spectrum for $t_1=t_2=\Delta_{tr}=0$. (c,d) Variation of the RIXS spectrum up to $\hbar\omega<2~$eV as a function of trigonal distortion, $\Delta_{tr}$ for the case of (c) zero coupling ($t_1=t_2=0$) (d) weak coupling ($t_1=0.1$~eV and $t_2=0$) between Ru ions. The RIXS intensity is shown on a linear color scale. The energy transfer, $\hbar\omega$, and trigonal distortion, $\Delta_{tr}$, are shown on the vertical and horizontal axes, respectively. \textbf{(e-i):} (e) Variation of the RIXS spectrum as a function of $t_1$ with $t_2=\Delta_{tr}=0$. (f) is the same as (e) but zoomed into the region with $\hbar\omega<0.3~$eV marked by the dotted yellow box. The excitation energies are plotted as a function of $t_1$ on top of the RIXS spectra in (f) as white dotted lines. The solid yellow line in (f) highlights the evolution of the $(L=0,S=\frac{3}{2},J=\frac{3}{2})\bigotimes (L=1,S=1,J=1)$ multiplet as a function of $t_1$. (g) Variation of the RIXS spectrum as a function of $\Delta_{tr}$ for fixed $t_1=0.75~$eV and $t_2=0.05$. For all calculations in (a-f), $U, J_H$ and $\lambda$ are fixed at $U=2.5~$eV, $J_H=0.3$~eV and $\lambda=0.15~$eV. (h,i) Comparison between simulated RIXS spectra using (h) $\Delta_{tr}=-0.1$~eV (\textbf{I}) and (i) $\Delta_{tr}=0.14$~eV (\textbf{II}) with the measured intra-$t_{2g}$ RIXS spectrum of Ba$_3$LaRu$_2$O$_9$ and Ba$_3$YRu$_2$O$_9$/Ba$_3$InRu$_2$O$_9$, respectively. The two $\Delta_{tr}$ values have been marked as vertical dashed lines in (g). To compare with experiment, the calculated spectra in (h) and (i) have been scaled by the same factor. The rest of the parameters are the same in (h) and (i) with $U=2.5~$eV, $J_H=0.3~$eV, $\lambda=0.15~$eV, $t_1=0.75~$eV, $t_2=0.05~$eV.}
\end{figure*}

To resolve this conundrum, we carried out RIXS modelling for the dimer by taking into account the six $t_{2g}$ orbitals on the two Ru ions (See Supplemental Materials \citep{supp} for details). Given the local $D_{3h}$ symmetry, there are six independent parameters within this minimal model\citep{LiyingPRB}: Hubbard interactions ($U$), Hund's coupling ($J_H$), SOC ($\lambda$), trigonal distortion ($\Delta_{tr}$) and hopping between singly degenerate $a_{1g}$ and doubly degenerate $e_{g}^\pi$ orbitals, denoted by $t_1$ and $t_2$ [See Fig.~\ref{structure}~(c)], respectively. To reduce the number of parameters, we fix the intra-atomic interactions to $U=$~2.5~eV, $J_H=$~0.3~eV, and $\lambda=$~0.15~eV, which show little variation across different Ru-compounds\citep{K2RuCl6RIXS,Ca2RuO4RIXS,BlairPRB2023}, and focus on the effects of $\Delta_{tr}$, $t_1$ and $t_2$. Due to the short Ru-Ru bond, $\Delta_{tr}$, $t_1$ and $t_2$ are expected to be substantially modified from their corresponding values with well-separated RuO$_6$ octahedra. In particular, a small separation between transition metal ions enables direct overlap between the axial $a_{1g}$ orbitals and can significantly increase $t_1$, which has been shown to be comparable or even larger than the intra-atomic interactions in some face-sharing Ir dimers\citep{Revelli2019,In_Irdimer}.

Qualitatively, the importance of $t_1$ in the present Ru dimer compounds is clear by comparing the measured RIXS spectra [Fig.~\ref{rixsdata}~(c)] to the prediction of a local picture by setting $t_1=t_2=0$ [Fig.~\ref{theory1}~(a-d)]. In this limit, the Ru$_2$O$_9$ dimer can be thought of as two weakly coupled Ru ions with a $t_{2g}^3$ and $t_{2g}^4$ electronic configuration, respectively, whose electronic states consist of well-separated Hund's multiplets as shown in Fig.~\ref{theory1}~(a). Due to the residual orbital degeneracy, the ground $(L=1,S=1)$ multiplet in the $t_{2g}^4$ case is further split by SOC into the $J=0, J=1$ and $J=2$ multiplets. As shown in Fig.~\ref{theory1}~(b), the RIXS spectrum in the localized limit consists of an intense low energy feature at $\hbar\omega\lesssim$0.3~eV and a weaker feature at much higher energy with $\hbar\omega\sim$0.7~eV-0.9~eV, corresponding to transitions within the ground state multiplet and to the excited Hund's multiplets, respectively [red and blue arrows in Fig.~\ref{theory1}~(a)]. Crucially, the large separation between these excitations, which is a direct consequence of the small $\lambda$ compared to $J_H$ in the ruthenates (in contrast to $\lambda\gtrsim J_H$ in iridates\cite{BoYuanRIXS}), implies that they are weakly coupled and hence very robust against small perturbations. This is confirmed by the simulated RIXS spectra in Fig.~\ref{theory1}~(c)-(d), where we show the variation of the RIXS spectrum up to $\hbar\omega<2~$eV (vertical axis) as a function of changes in local structure parametrized by $\Delta_{tr}$ (horizontal axis) for the case of (c) $t_1=t_2=$0 and (d) $t_1=0.1$~eV and $t_2=$0. Other than a slight shift in energy of different RIXS features, the RIXS spectrum varies smoothly with $\Delta_{tr}$ and resembles that of Fig.~\ref{theory1}~(b) for the entire range of $\Delta_{tr}$. Within a local picture, it is therefore difficult to explain the abrupt changes in RIXS spectrum observed with minimal changes in local structure across the Ba$_3$MRu$_2$O$_9$ series. Moreover, the RIXS spectrum of Ba$_3$LaRu$_2$O$_9$ shows only one peak at $\sim 0.7~$eV, completely different from the simulation in Fig.~\ref{theory1}(b)-(d) which always shows two peaks below 1~eV. As we show below, both of these observations can be explained by the presence of a large inter-site hopping $t_1$.

As shown by the calculated $t_1$-dependence of the excitation energies and the RIXS spectra in Fig.~\ref{theory1}~(e), increasing $t_1$ moves up a bulk of the low energy states within the $(L=0,S=\frac{3}{2})\bigotimes(L=1,S=1)$ manifold together with most of its spectral weight. This part of the spectral weight eventually merges with the high energy transitions for $t_1\gtrsim0.7~$eV, and provides a natural explanation for the presence of only one peak in the RIXS spectrum of Ba$_3$LaRu$_2$O$_9$.
 
More importantly, unlike the smooth variation of the RIXS spectrum with $\Delta_{tr}$ in the local limit [Fig.~\ref{theory1}~(c)-(d)], varying $\Delta_{tr}$ in the limit of large $t_1$ as shown in Fig.~\ref{theory1}~(g) leads to an abrupt change in the RIXS spectrum at $\Delta_{tr}^c\approx0.13~$eV, across an electronic transition where the first excited states cross the ground states [See Supplemental Materials for justification of the choice of $t_1$ and $t_2$ in Fig.~\ref{theory1}~(g)-(i)]. Remarkably, as shown by the direct comparison between theory and experiment in Fig.~\ref{theory1}~(h)-(i), the simulated RIXS spectrum at $\Delta<\Delta_{tr}^c$ (\textbf{I}) and $\Delta>\Delta_{tr}^c$ (\textbf{II}) provides an excellent description of the spectrum of Ba$_3$LaRu$_2$O$_9$ and Ba$_3$InRu$_2$O$_9$/Ba$_3$YRu$_2$O$_9$, respectively, by capturing the suppression/enhancement of the high/low energy peaks. It is worth emphasizing that the abrupt changes in the RIXS spectrum at $\Delta_{tr}^c$ [Fig.~\ref{theory1}~(g)-(i)] come from a modification of the RIXS matrix elements due to the abrupt change in ground-state wave-function. Consequently, only a small tweak in $\Delta_{tr}$ is needed, consistent with a very small change in local structure. Our model also predicts a large drop in effective moment from $\sim3.7~\mu_B$ to $\sim 1~\mu_B$ across the electronic transition \citep{supp}. These values are in reasonable agreement with the values from magnetic susceptibility measurements \citep{supp,PhysRevB.95.184424,PhysRevMaterials.4.064409} and provide further support for our theory. 
 
The ability to drive an electronic transition at very small $\Delta_{tr}^c$ [Fig.~\ref{theory1}~(g)] is a direct manifestation of the enhanced electronic instability of the dimer in the limit of large $t_1$ compared to the local limit [Fig.~\ref{theory1}~(c)-(d)]. This can be understood from the evolution of the low energy states as a function of $t_1$ shown in Fig.~\ref{theory1}~(f). Unlike the local model where the first excited states occur at $\sim\frac{2}{3}\lambda\sim0.1~$eV, there is an abundance of low energy states with $\hbar\omega<50~$meV at large $t_1$, which could cross the ground states and drive an electronic transition in the dimer with a small change in $\Delta_{tr}$. Further demonstrating the enhanced electronic instability of the dimer, we found that an electronic transition can also be induced by other small perturbations such as $t_2$ \citep{supp}. However, the predicted change in RIXS spectrum by varying $t_2$ is inconsistent with the experimental observations. We therefore attribute the observed change in RIXS spectrum to a change in $\Delta_{tr}$. 

Of special importance among the low energy states is the doublet originating from the excited $(L=0,S=\frac{3}{2},J=\frac{3}{2})\bigotimes (L=1,S=1,J=1)$ multiplet [marked by the yellow solid line in Fig.~\ref{theory1}~(f)] corresponding to a scenario where $t_{2g}^3$ is in its ground state and $t_{2g}^4$ is in the excited $J=1$ state in the $t_1=0$ limit. Due to the reduced SOC in Ru, this doublet is pushed to a very small energy ($\sim~30~$meV) in the limit of large $t_1$ and strongly hybridizes with states from the ground multiplet. As illustrated in the Supplemental Materials\citep{supp}, such hybridization is crucial for the electronic transition shown in Fig.~\ref{theory1}~(g), and marks a clear distinction between Ir and Ru dimer systems. The relevance of the $J=1$ excited multiplet is reminiscent of the famous excitonic magnetism proposed for $t_{2g}^4$ ions\citep{excitonic}, which is difficult to realize in isolated Ru$^{4+}$ and Ir$^{5+}$ ions with a large spin-orbit gap. Our theory therefore suggests an interesting mechanism for its realization in MO systems, where the energy of the J=1 exciton can be substantially lowered by large inter-site hopping. Although not strictly applicable to the present system with 7 electrons per dimer, Ru-dimer compounds Ba$_3$MRu$_2$O$_9$ with M=Ce$^{4+}$, Pr$^{3+}$ \citep{doi:10.1021/jacs.9b03389, 10.1246/bcsj.76.1093} and 8 electrons per Ru$_2$O$_9$ dimer might be promising candidates for a MO version of excitonic magnetism.

In conclusion, by carrying out Ru M$_3$-edge RIXS measurements on a well-known series of Ru dimer compounds with the 6H-perovskite structure, Ba$_3$MRu$_2$O$_9$ (M$^{3+}$=In$^{3+}$, Y$^{3+}$, La$^{3+}$), we find direct spectroscopic evidence for an extremely fragile molecular orbital state, manifested as an abrupt change in RIXS spectrum concomitant with a very subtle change in local structure tuned by the M-site ion. By modelling our RIXS spectra, we attribute the enhanced electronic instability in these dimer compounds to a cooperative effect between large inter-site hopping and small SOC, a unique combination found only in the Ru but not Ir MO systems. Our study therefore highlights the Ru MO systems as ideal platforms for studying quantum phase transitions involving molecular orbitals.  

\begin{acknowledgments}
BY acknowledges support from the NSERC postdoctoral fellowship. JPC is supported by the NSERC Discovery Grant program. BHK was supported from the Institute for Basic Science in the Republic of Korea through the Project No. IBS-R024-D1. This research used the SIX beamline (2-ID) of the National Synchrotron Light Source II, a U.S. Department of Energy (DOE) Office of Science User Facility operated for the DOE Office of Science by Brookhaven National Laboratory under Contract No. DE-SC0012704.
\end{acknowledgments}

\end{document}